\title{On the concentration of eigenvalues of random symmetric matrices
}
\author{
Michael Krivelevich\thanks{
Department of Mathematics, Raymond and Beverly Sackler Faculty of Exact
Sciences, Tel Aviv University, Tel Aviv 69978, Israel.
Email: krivelev@math.tau.ac.il.
}
\and Van H. Vu
\thanks{
Microsoft Research, 1 Microsoft Way, Redmond, WA 98052, USA.
E-mail: vanhavu@microsoft.com.
}
     }
\date{}
\documentclass [11pt]{article}
\usepackage{latexsym}
\newtheorem{theorem}{Theorem}
\newtheorem{theo}{Theorem}[section]

\newtheorem{lemma}[theo]{Lemma}
\newtheorem{coro}[theo]{Corollary}

\newcommand{\Proof}{\noindent{\bf Proof.}\ \ }
\newcommand{\CA}{{\cal A}}
\newcommand{\CB}{{\cal B}}

\newcommand{\bv}{{\bf v}}
\newcommand{\bu}{{\bf u}}
\newcommand{\bw}{{\bf w}}
\newcommand{\bx}{{\bf x}}
\newcommand{\byy}{{\bf y}}
\newcommand{\bJ}{{\bf 1}}
\newcommand{\bZ}{{\bf 0}}
\newcommand{\Ex}{{E}}  % {{\Bbb E}}
\newcommand{\Rx}{{R}}  % {{\Bbb R}}
\newcommand{\ep}{{\epsilon}}

\begin{document}
\maketitle
%\setcounter{page}{0}
%\centerline{Technical Report Number }

%\centerline{MSR-TR-2000-60}
\begin{abstract}
We prove that the few largest (and most important) eigenvalues of
 random symmetric matrices of various kinds are very strongly
 concentrated. This strong concentration enables us to compute
 the means of these eigenvalues with high precision. Our approach uses
 Talagrand's inequality and is very different from standard
 approaches.

\end{abstract}

%\newpage
\section{Introduction}

In this paper we consider the eigenvalues of random symmetric matrices
whose diagonal and upper diagonal
entries are  independent random variables. Our goal is to study
few largest/smallest  eigenvalues of such a matrix.
Let us begin with a
version of Wigner's famous semi-circle law \cite{Wig}, due to
Arnold \cite{Arn, Meh}, which describes the limiting behavior of
the bulk of the spectrum of a random matrix of this type.

\vskip 3mm

\noindent{\bf Semi-circle law.}
{\em
 For $ 1\le i \le j \le n$ let $a_{ij}$
be real value random variables such that all $a_{ij}$, $i < j$
have the same distribution and all $a_{ii}$ have the same
distribution. Assume that all central moments of the $a_{ij}$ are
finite and put $\sigma^2 =\sigma^2 (a_{ij})$. For $i <j$ set
$a_{ji}=a_{ij}$ and let $A_n$ denote the random matrix
$(a_{ij})^{n}_1$. Finally, denote by $W_n (x)$ the number of
eigenvalues of $A_n$ not larger than $x$, divided by $n$. Then
$$
\lim _{n \rightarrow \infty} W_n(x 2\sigma \sqrt{n}) = W(x)\ ,
$$
in distribution, where $W(x)=0$ if $x \le -1$,$ W(x)=1$
if $x \ge 1$ and $W(x)= \frac{2}{\pi} \int_{-1}^x (1-x^2)^{1/2}
dx$ if $ -1 \le x \le 1$.
}

\vskip 3mm

The semi-circle law gives only a limit distribution and  does not
tell anything about the behavior of the largest/smallest (and usually
 most important) eigenvalues. These eigenvalues were studied in several
papers \cite{Juh, FK,SS, So}. The method used in these papers is
to estimate the expectation of the trace of a high power of the
matrix. This frequently leads to a sharp upper bound on the
largest eigenvalue (see Section 2).

Given a symmetric matrix $A$, we denote by
$\delta_1 (A) \ge \delta _2(A) \ge \dots \ge \delta_n (A)$ the
eigenvalues of $A$. Furthermore,
let $\lambda_1 (A) = \max_{i=1}^n (|\delta_i A)|) =\max (|\delta_1
(A)|, |\delta_n (A)|)$ and $\lambda_2 (A) = \max(|\delta_2 (A)|,
|\delta_n (A)|)$.

The purpose of this paper is to prove large deviation bounds for
$\lambda_1, \lambda_2, \delta_1, \delta_2$ and $\delta_n$. We
believe that these results are of interest for a number of
reasons. The first is that our results are obtained under a very
general assumption on the distribution of the entries of a random
symmetric matrix. Secondly, our large deviation bounds turn out to
be very strong. Moreover, they are sharp, up to a constant in the
exponent, in a certain deviation range. Also, our method appears
to be new; it makes a novel application of the recent and powerful
inequality of Talagrand \cite{Tal}. Finally, since bounds on the
largest eigenvalues of  a symmetric random matrix are widely used
in many applications in Combinatorics and Theoretical Computer
Science, we believe that our results have a potential in these
areas.  As an example of such an application, we would like to
mention a paper \cite{KV} of the present authors, where  a version
of our theorems has been used to design approximation algorithms
with expected polynomial running time for such important
computational problems as finding the chromatic number and the
independence number of a graph.

 Our first result involves the following  general model.
  Let $a_{ij}$ ($1 \le i \le j \le n$) be independent random
variables, with absolute value at most  1. A symmetric random
matrix $A$ is obtained by  defining $a_{ji} = a_{ij}$ for all $i <
j$.

\begin{theorem}\label{th1}
There are positive constants $c$ and $K$ such that for any $t > K$,
$$
Pr[|\lambda_1 (A) -\Ex (\lambda_1 (A))| \ge t ] \le
e^{ -ct^2 }\ .
$$
The same  result holds for both $\delta_1 (A)$ and $\delta_n (A)$.
\end{theorem}

The bound in Theorem \ref{th1} is sharp, up to the constant $c$, when
$t$ is sufficiently large. The surprising fact about this theorem
is that it requires basically no knowledge about the distributions of
the $a_{ij}$.
% This gives us a strong reason to believe that this
%theorem cannot be obtained by the standard moment method, which
%usually requires some strong assumption about the distributions.

Our second theorem provides a large deviation result for the second
largest eigenvalue of a symmetric random matrix $A$, under the
additional assumption that all non-diagonal entries of $A$ have the
same expectation $p>0$.

\begin{theorem}\label{th2}
For every constant $p>0$ there exists constants $c_p,K_p>0$ so
that the following holds. If in addition to the conditions of
Theorem \ref{th1}, the random variables $a_{ij}$, $1\le i< j\le n$
satisfy $\Ex[a_{ij}]=p$, then for all $t>K_p$, $$
Pr[\lambda_2(A)-\Ex(\lambda_2(A))|\ge t]\le e^{-c_pt^2}\ . $$ The
same result holds for $\delta_2(A)$.
\end{theorem}

One particular application of the above theorem arises when all
diagonal entries of $A$ are 0, and each non-diagonal entry of $A$
is a Bernoulli random variable with parameter $p$, i.e.
$Pr[a_{ij}=1]=p$, $Pr[a_{ij}=0]=1-p$. In this case $A$ can be
viewed as the adjacency matrix of the {\em random graph} $G(n,p)$.
Thus Theorem \ref{th2} provides in this case a large deviation
result for the second eigenvalue of a random graph. In fact, for
this special case, Theorem \ref{th2} can be extended for $p$
decreasing in $n$ (see Section 5).

The rest of the paper is organized as follows. In the next
section, we collect some information about the expectations of the
eigenvalues in concern. More interesting, it turns out that  our
theorems can sometimes be used to estimate these expectations. The
proofs of Theorems \ref{th1} and \ref{th2} appear in Sections 3 and 4,
respectively. We end with Section 5, which contains few remarks
and open questions.

In what follows, a matrix is always symmetric and of
order $n$, if not otherwise specified. We assume that $n$ tends to
infinity and the asymptotic notations (such as $o$, $O$, etc) are
understood under this assumption. The letter $c$ denotes a
positive constant, whose value may vary in different  occurrences.
Bold lower case letters such as $\bx, \byy$ denote vectors in
$\Rx^n$ and $\bx \byy$ is the inner product of $\bx$ and
$\byy$. Given a matrix $A$, $\bx A \byy$ is the inner product of
$\bx$ and $A\byy$. $\bJ$ is the all one vector.

\section{Expectations}\label{exp}

In this section, we present several results about the expectation
of the relevant eigenvalues. We also show that our theorems can
be used to determine these expectations in some cases.

 Let $a_{ij}$, $i \le j$ be independent random variables bounded
in their absolute values by 1. Assume that for $i <j$, the $a_{ij}$
have common expectation $p$ and variance $\sigma^2$. Furthermore, assume
that $\Ex[a_{ii}]= \nu$ for all $i$. F\"uredi and Koml\'os (\cite{FK},
Theorem 1), showed that if $p > 0$ then

\begin{equation}\label{fk1}
\Ex[\lambda_1 (A)] = (n-1)p + \nu  + \sigma^2/p + o(1)\ .
\end{equation}
Also, in this case under a weaker assumption $VAR[a_{ij}]\leq \sigma^2$
for all $1\le i\le j\leq n$ the argument of F\"uredi and Koml\'os gives:
\begin{equation}\label{fk2}
\Ex [\lambda_2 (A)] \le 2\sigma\sqrt{n} + O (n^{1/3} \log n)\ .
\end{equation}

The situation changes when $p=0$. In the same paper, F\"uredi
and Koml\'os (implicitly) showed that in this case (again assuming only
$VAR[a_{ij}]\leq \sigma^2$)
\begin{equation}\label{fk3}
 \Ex[\lambda_1(A)] \le 2 \sigma \sqrt{n} + O(n^{1/3} \log n)\ .
\end{equation}

F\"uredi and Koml\'os also claimed that if $Var[a_{ij}]= \sigma^2$
then  with probability tending to 1, $\lambda_1 (A) \ge 2 \sigma
\sqrt{n} +O( n^{1/3} \log n)$.
%However, their proof of this (lower
%bound) statement \cite{FK} is incomplete.

%In \cite{SS}, Sinai and Soshnikov showed that if $a_{ij}$ have
%symmetric distribution, then $Pr[\lambda_1(A)\le 2 \sigma
%\sqrt{n}+o(1)]=1-o(1)$. They also claimed that a similar result
%would hold without the symmetric assumption. However, no proof was
%provided. Furthermore, Soshnikov proved in [So] that if the
%distributions of $a_{ij}$ are symmetric, then for every fixed $k$,
%asymptotically almost surely the first $k$ eigenvalues of $A$ can
%be bounded from above by $2\sigma\sqrt{n}+o(1)$. The proof of this
%result and that of (3) relies on a trace estimate and do not seem
%to give a lower bound for the eigenvalues in question.

Using  our Theorem \ref{th1}, we first show that a statement
slightly weaker than (\ref{fk1}) holds under a more general
assumption that the variances are upper bounded by $\sigma^2$, but
are not necessarily equal. Next, we prove a lower bound stronger
than that stated by  F\"uredi and Koml\'os.

\begin{coro}\label{co1}

If all entries $a_{ij}$ of the random symmetric matrix
$A=(a_{ij})$ are bounded in absolute value by 1, and all
non-diagonal entries have common expectation $p>0$, then $$
\Ex[\lambda_1(A)]=np+O(\sqrt{n})\ . $$

\end{coro}

\Proof For each entry $a_{ij}$, one can define a random variable
$c_{ij}$, satisfying $|c_{ij}|\leq 1$, $\Ex[c_{ij}]=0$,
$VAR[c_{ij}]=1-VAR[a_{ij}]$. Let now $b_{ij}=a_{ij}-c_{ij}$. Then
clearly $\Ex[b_{ij}]=p$, $VAR[b_{ij}]=1$. Denote $B=(b_{ij})$,
$C=c_{ij}$, then $A=B+C$. Hence $\lambda_1(A)\leq
\lambda_1(B)+\lambda_1(C)$. Applying (\ref{fk1}), (\ref{fk3}) and
Theorem \ref{th1}, we obtain:
\begin{eqnarray*}
Pr[\lambda_1(B)\le np +O(1)] &\ge& \frac{3}{4}\,,\\
Pr[\lambda_1(C)\le O(\sqrt{n})] &\ge& \frac{3}{4}\,,
\end{eqnarray*}
and thus $Pr[\lambda_1(A)=np+O(\sqrt{n})]\ge 1/2$. Invoking
Theorem \ref{th1} once again, we get the desired result.
\quad\quad $\Box$

By the same argument, one can show that if $\sigma= \omega
(n^{-1/2})$, then $E(\delta_n)= 2 \sigma n^{-1/2} (1 + o(1))$.

\begin{coro}\label{co2}

If all entries $a_{ij}$ of the random symmetric matrix
$A=(a_{ij})$ have common expectation 0 and variance $\sigma^2$,
then

$$ E[\lambda_1(A)] \ge 2 \sigma n^{1/2} + O(\log^{1/2} n ) \ . $$

\noindent Consequently, with probability tending to 1,

$$\lambda_1 (A) \ge 2 \sigma n^{1/2} + O(\log^{1/2} n ) \ . $$

\end{coro}

\Proof For the sake of simplicity, we assume $\sigma=1/2$.
Furthermore, set $\mu= n^{1/2}$, $k= \lceil \mu \log^{1/2} n
\rceil$ and $x= a \log^{1/2} n$, where $a$ is a positive constant
chosen so that the following two inequalities hold:

\begin{equation}\label{muk1}
\mu^k/ k^{5/2} \ge 2 (\mu -x/2)^k
\end{equation}

\begin{equation}\label{muk2}
\sum_{t= \frac{a}{2} \log^{1/2} n}^{\infty}  e^{2t \log^{1/2}n - c
t^2} = o(1),
\end{equation}

\noindent where $c$ is the constant in Theorem 1. Without loss of
generality, we assume that $k$ is  an even  integer and let $X$ be
the trace of $A^k$. It is trivial that $E[X] \le n
E[\lambda_1^k]$. On the other hand, a simple counting argument
(see \cite{FK}) shows that

$$E[X] \ge \frac{1}{(k/2) +1} {k \choose {k/2}} \sigma^k
n(n-1)\dots(n-(k/2)). $$

It follows that
 \begin{equation}\label{muk3}
E[\lambda_1^k] \ge \frac{1}{(k/2) +1} {k \choose {k/2}}\sigma^k
(n-1)\dots(n-(k/2)) \ge \mu^k/ k^{5/2}\,.
 \end{equation}

Assume, for contradiction, that $E(\lambda_1) \le \mu -x$. It
follows from this assumption that

\begin{equation}\label{muk4}
E[\lambda_1^k ] \le (\mu-x/2)^k + \sum_{t=x/2}^{\infty}
\bigl(\mu-x +(t+1) \bigr)^k  Pr[\lambda_1 \ge \mu -x + t]\,.
\end{equation}

By Theorem \ref{th1}, $ Pr(\lambda_1 \ge \mu -x + t) \le e^{-ct^2}$ for
all $t \ge x/2$.
% Moreover, $\bigl(\mu-x +(t+1) \bigr)^k -
%\bigl(\mu-x + t \bigr)^k \le k (\mu -x + (t+1))^{k-1} \le
%2(\log^{1/2} n)  (\mu -x + (t+1))^k $.
Thus
(\ref{muk1}),(\ref{muk3}) and (\ref{muk4}) imply

\begin{equation}\label{muk5}
\sum_{t=x/2}^{\infty}  (\mu -x + (t+1))^k e^{-ct^2} \ge \mu^k/
k^{5/2} - (\mu-x/2)^k \ge (\mu -x/2)^k.
\end{equation}

Since $(\mu -x + (t+1))^k /(\mu-x/2)^k \le e^{(1+o(1)) t k/ \mu}
=e^{ (1+o(1)) t\log^{1/2} n }$, (\ref{muk2}) and (\ref{muk5})
imply a contradiction, and  this completes the proof. \quad \quad
$\Box$

 To end this section, let us mention few recent results of
Sinai and Soshnikov. In \cite{SS}, Sinai and Soshnikov showed that
if $a_{ij}$ have symmetric distributions and their moments satisfy
some mild assumptions, then $Pr[\lambda_1(A)\le 2 \sigma
\sqrt{n}+o(1)]=1-o(1)$. They also stated that a similar result
would hold without the symmetric assumption.
%However, no proof was provided.
Furthermore, Soshnikov proved in [So] that under the same
assumptions about $a_{ij}$, the joint distribution of the $k$
dimensional  random vector formed by the first $k$ eigenvalues,
scaled properly,  tends to a weak limit, for any fixed $k$.

\section{Proof of Theorem \ref{th1}}

The key tool of the proof is a powerful  concentration result, due
to Talagrand \cite{Tal}. To state this inequality, we first need
to define the so-called Talagrand distance in a product space. Let
$\Omega_1,\ldots,\Omega_m$ be probability spaces, and let $\Omega$
denote their product space. Fix a set $\CA \subset \Omega$ and a
point $\bx =(x_1, \dots, x_m) \in \Omega$. We say that $x$ has
Talagrand distance $t$ from $\CA$ if $t$ is the smallest number
such that the following holds. For any real vector ${\bf \alpha} =
(\alpha_1, \dots, \alpha _m),$ there is a point $\byy = (y_1,
\dots y_m) \in \CA$ such that

$$\sum_{x_i \neq y_i}|\alpha_i| \le t\left(\sum_{i=1}^n
\alpha_i^2\right)^{1/2}. $$

Let $\CA_t$ denote the set of all points with Talagrand distance
at most $t$ from $\CA$.  Talagrand proved that for any $t \ge 0$,

$$ \Pr[\CA]Pr[\overline{ \CA_t}] \le e^{-t^2/4},  $$

\noindent where $\overline{ \CA_t}$ denotes the complement of
$\CA_t$. Remarkably, the rather abstract and difficult definition
of the Talagrand distance suits our problem perfectly, as shown in the
proof below.

\vskip 2mm

 Consider the product space spanned
by $a_{ij}, 1 \le i \le j \le n$. A vector in this space
corresponds to a random matrix. Let $m$ be a median of
$\lambda_1$ and let $\CA$ be the set of all matrices (vectors) $T$
such that $\lambda_1 (A) \le m$. By definition, $Pr [\CA] \ge 1/2$. By a
well known fact in linear algebra
$$
\lambda_1 (A) = \max_{\| \bv\|=\|\bw\| =1}
\sum_{ 1 \le i < j \le n} (v_iw_j + v_j w_i) t_{ij} +
\sum_{i=1}^n v_iw_i t_{ii}\ .
 $$

Consider a matrix $X=(x_{ij})$
where $\lambda_1 (X) \ge m + t$. There are
vectors  $\bv=(v_1, \dots, v_n)$, $\bw=(w_1, \dots, w_n)$ with
norm 1 such that
$$
\bv X \bw=  \sum_{ 1 \le i \le j \le n} (v_iw_j + v_j w_i)
x_{ij} + \sum_{i=1}^n v_iw_i x_{ii} \ge m  + t\ .
$$
On the other hand, for any $Y=(y_{ij}) \in \CA$
$$
\bv Y\bw =\sum_{ 1 \le i \le j \le n} (v_i w_j + v_j w_i)  y_{ij} +
\sum_{i=1}^n v_i w_i y_{ii} \le m\ .
$$
Set $\alpha _{ij} = (v_i w_j + v_j w_i) $  for $ 1 \le i <j \le n$
and $\alpha_{ii} = v_i w_i$ for $  1 \le i \le n$. It is easy to show
that
$$
\sum_{1 \le i \le j \le n} \alpha_{ij}^2 < 2(\sum_{1 \le i \le
n} v_i^2)(\sum_{i=1}^n w_i^2) =2\ .
$$
Since $ |x_{ij} - y_{ij}| \le 2$, we have
$$
\sum_{x_{ij } \neq y_{ij} } |\alpha_{ij}| \ge t/2 >
\frac{t}{\sqrt{8}} \left(\sum_{1 \le i \le j \le n} \alpha_{ij}^2
\right)^{1/2}\ .
$$

\noindent By definition, it follows that $X \in \overline
{\CA_{t/\sqrt{8}} }$. Therefore, by Talagrand's inequality
\begin{equation}\label{lb31}
Pr[\lambda_1 (A) \ge m + t] \le 2 e^{-t^2/ 32}\ .
\end{equation}

\noindent Let $\CB$ be the set of $A$ such that $\lambda_1 (A) \le
m-t$. By a similar argument, one can show that if $\lambda_1 (A)
\ge m$ then $A \in \overline {\CB_{t/\sqrt{8}}}$. Recall that
$Pr[\lambda_1(A)\geq m]\geq 1/2$. Thus Talagrand's
inequality implies
\begin{equation}\label{lb32}
Pr[\lambda_1 (A) \le m - t] \le 2 e^{-t^2/ 32}\ .
\end{equation}
>From here, one can derive that the difference between
the median and the expectation of $\lambda_1$ is bounded by a
constant:
\begin{eqnarray}
|\Ex (\lambda_1 (A)) - m| &\le& \Ex (|\lambda_1 - m|) \le
\int_{0}^{\infty} t Pr[|\lambda_1 (A) -m| \ge t] dt\nonumber \\
 &\le&
\int_0^{\infty} 4te^{-t^2/32} dt = 64\ .\label{lb33}
\end{eqnarray}
Inequalities (\ref{lb31}), (\ref{lb32}) and (\ref{lb33}) together imply
the desired deviation bound
for $\lambda_1 (A)$. The statements involving $\delta_1(A)$ and
$\delta_n(A)$ can be proved in a similar way, using the following
equalities:
$$
\delta_1 (A)= \max_{\bx, \|\bx\|=1} \bx A \bx.
$$

$$
\delta_n (A) = \min_{\bx, \|\bx\|=1} \bx A\bx\ .
$$

\vskip 3mm

\noindent{\bf The sharpness of the result.} The following example shows
that the bound in Theorem \ref{th1}  is best possible, up to a
multiplicative constant in the exponent.

Assume that $a_{ij}$, $1\le i \le j \le n$,  have the following
distribution: $a_{ij}=1$ with probability $p$ and $a_{ij}=-p/q$
with probability $q=1-p$. A matrix is {\it fat} if it contains an
all 1 principle sub-matrix of size $  \Ex [\lambda_1]+t$. It is
trivial that if $A$ is fat then $\lambda_1(A) \ge \Ex[\lambda_1 ]
+ t$. On the other hand, the probability that a matrix is fat is
at least $p^{(\Ex[\lambda_1] + t)^2}= e^{-\bigl[\Ex(\lambda_1 ] +
t \bigr)^2 \log \frac{1}{p} }$. Thus, if $p$ is a positive
constant and $t$ is of order $\Omega (\Ex[\lambda_1])$, then
$$
Pr [|\lambda_1  -\Ex [\lambda_1]| \ge t] \ge e^{-c t^2},
$$
for some positive constant $c$.

\section{Proof of Theorem \ref{th2}}
 Given a symmetric matrix $A$, $\lambda_2(A)$ can be expressed as
follows \cite{Gan}:
$$
\lambda_2 (A) =\min_{\bZ\ne\bv \in \Rx^n} \max_{ {{\bx,\byy} \atop
{\|\bx\|=\|\byy\|=1}} \atop {\bx\bv=\byy\bv=0}} \bx A\byy\ .
$$

Define
$$
\mu_2(A) = \max_{{      {\bx,\byy}
                 \atop {\bx\bJ=\byy\bJ=0}}
                 \atop {\|\bx\|=\|\byy\|=1}}
           \bx A\byy\,.
$$
It is clear that $\mu_2 (A) \ge \lambda_2 (A)$ for any matrix $A$.
In the rest of  the proof, we use shorthands $\mu_2, \lambda_2$ for
$\mu_2(A), \lambda_2 (A)$, respectively, where $A$ is distributed as
described in the theorem formulation. Similar to the previous
section, by Talagrand's inequality we can show

\begin{lemma}\label{le41}
There are positive constants $c$ and $K$ such that for any $t>K$
$$
Pr [|\mu_2 - \Ex (\mu_2)| \ge t ] \le e^{-c t^2}\ .
$$
\end{lemma}

Set  $A'= A-pJ_n$, where  $J_n$ denotes the all one matrix of
order $n$. It is easy to show that $\mu_2 (A) \le \lambda_1 (A')$.
Indeed,
\begin{eqnarray*}
\mu_2&=&\max_{{{\bx,\byy}\atop{\bx\bJ=\byy\bJ=0}}
              \atop{\|bx\|=\|\byy\|=1}}
      \bx(A'+pJ_n)\byy
      = \max_{{{\bx,\byy}\atop{\bx\bJ=\byy\bJ=0}}
              \atop{\|bx\|=\|\byy\|=1}}
      \bx A'\byy\\
&\leq& \max_{\|bx\|=\|\byy\|=1}\bx A'\byy=\lambda_1(A')\ ,
\end{eqnarray*}
where the second equality uses the fact that $\bx$ and $\byy$ are
orthogonal to the vector of all 1's and are thus orthogonal to every
row of $J_n$.

Since each non-diagonal entry of $A'$ has mean 0 and is bounded in
absolute value by $1+p\leq 2$, by the result (\ref{fk3}) of F\"uredi
and Koml\'os,  $\Ex[\lambda_1 (A')] \le 3 \sqrt{n}$.
Assume that $t \ge 10 \sqrt{n}$; Theorem \ref{th1} implies then
$$
Pr [ |\lambda_2 -\Ex(\lambda_2)| \ge t ] \le
Pr [ \lambda_1 (A') \ge \Ex (\lambda_1 (A')) + t/2 ] \le
e^{-ct^2}\ .
$$

The proof of the  case  $t < 10\sqrt{n}$ is harder and is based
on the following two lemmas.

\begin{lemma}\label{le42}
For every constant $p>0$ there exist constants $c_p>0$, $K_p>0$ so that
for any $K_p<t<10\sqrt{n}$, there is a positive
number  $\ep_t =O(t^{1/2} (np)^{-1/2})$ such that
$$
Pr [\mu_2-(1+ \ep_t)  \lambda_2  \ge t] \le e^{-c_p t^2}\ .
$$
\end{lemma}

\begin{lemma}\label{le43}
For every constant $p>0$ there exists a constant $L_p>0$ such that
$$
\Ex [\mu_2] -\Ex [\lambda_2 ] \le L_p\ .
$$
\end{lemma}

Assuming these two lemmas hold, we can finish the proof as
follows. First assume, without loss of generality, that $t \ge
5L_p$. Consider the upper tail:
\begin{eqnarray*}
Pr [\lambda_2 \ge \Ex(\lambda_2 ) + t ]  &\le&
Pr [\mu_2 \ge \Ex(\lambda_2) +t ] \\
&\le& Pr [\mu_2 \ge \Ex (\mu_2) + (t-L_p) ] \\
&\le& e^{- \Omega ( (t-L_p)^2) } = e^{-c_pt^2}\ ,
\end{eqnarray*}
by Lemma \ref{le41}.

Now consider the lower tail:
\begin{eqnarray*}
Pr [\lambda_2 \le \Ex[\lambda_2 ] - t ] &\le&
Pr [(1+\ep_t )\lambda_2 \le (1+ \ep_t) \Ex[\lambda_2 ] - t ]\\
&\le& Pr [\mu_2 \le  (1+ \ep_t) \Ex[\lambda_2 ] - t/2 ] +
Pr [\mu_2 -(1+ \ep_t) \lambda_2 \ge t/2 ] \ .
\end{eqnarray*}
By Lemma \ref{le42}
$$
 Pr [\mu_2 -(1+ \ep_t)\lambda_2 \ge t/2] \le e^{-c_pt^2}\,.
$$
On the other hand,
$$
 Pr [\mu_2 \le  (1+ \ep_t) \Ex[\lambda_2 ] - t/2 ] \le
\Pr [\mu_2 \le (1+\ep_t)\Ex[\mu_2] - t/2]\ .
$$
Given that $t$ is sufficiently large, $\ep_t
\Ex[\mu_t] = O(t^{1/2} ) \le t/4$. So, by Lemma \ref{le41},  the last
probability can also be bounded by $e^{-c_pt^2}$ and this completes
the proof. \quad\quad $\Box$

\vskip 3mm

To prove Lemmas \ref{le42} and \ref{le43}  we need three other lemmas.
The first two (Lemmas \ref{le44} and \ref{le45}) are linear algebraic
statements. The last one (Lemma \ref{le46}) is a statement about the
concentration of a certain random variable, which is a function of the
entries $a_{ij}$ of the random symmetric matrix $A$.

\begin{lemma}\label{le44}
Let $A$ be an $n$ by $n$ real symmetric matrix. Let $a$ satisfy $0\le
a<\sqrt{n}$.
Denote by  $\bv_1$ a unit eigenvector corresponding to $\lambda_1(A)$.
Assume there is a number $c_1, 0<|c_1| \le \sqrt{n}$
such that $\|\bJ - c_1 \bv_1 \| \le a$. Then
$$
\mu_2 (A) - \lambda_2 (A) \le \frac{2 a \lambda_2 (A)}{ \sqrt{n}-a}
+ \frac{a^2\lambda_1(A)}{(\sqrt{n}-a)^2}\ .
$$
\end{lemma}

\Proof
Note first that
$$
\|c_1\bv_1\|=\|(c_1\bv_1-\bJ)+\bJ\|\ge \|\bJ\|-\|c_1\bv_1-1\|
\ge \sqrt{n}-a\,.
$$
Assume that $\mu_2 (A) = \bx A\byy$, where $\bx,\byy$ are unit vectors
perpendicular to $\bJ$. Then
$$
\bx(c_1\bv_1)=\bx,(c_1\bv_1-\bJ+\bJ)=\bx(c_1\bv_1-1)
\le \|\bx\|\cdot\|c_1\bv_1-\bJ\|\le a\,.
$$
Notice that as $\|\bJ\|=\sqrt{n}y$ and $a<\sqrt{n}$, we have
$c_1\ne 0$. Define
$$
\bx'=x-\frac{\bx(c_1\bv_1)}{(c_1\bv_1)(c_1\bv_1)}\,c_1\bv_1\,.
$$
Then $\bx'$ is orthogonal to $\bJ$ and satisfies
$\|\bx'\|\le \|\bx\|=1$. Set $\bu=\bx-\bx'$. Then
$$
\|\bu\|=\frac{|\bx(c_1\bv_1)|}{\|c_1\bv_1\|}
       \leq \frac{a}{\sqrt{n}-a}\ .
$$
Similarly, set
$$
\byy'=\byy-\frac{\byy(c_1\bv_1)}{(c_1\bv_1)(c_1\bv_1)}\,c_1\bv_1\,,
$$
then $\byy'$ is a vector of norm at most 1 orthogonal to $\bv_1$.
Denoting $\bw=\byy-\byy'$, we can prove that
$\|\bw\|\le a/(\sqrt{n}-a)$.

By definition, $\lambda_2 (A) \ge | \bx'A \byy' |$. On the other
hand, by the Cauchy--Schwartz inequality
\begin{eqnarray*}
\bx A\byy- \bx 'A\byy' &=& (\bx'+\bu)A(\byy'+\bw)-\bx'A\byy'=
\bw A\bx'+\bu A\byy'+\bu A\bw\\
&\le& \|\bw\|\|A\bx'\|+\|u\|\|A\byy'\|+\lambda_1(A)\|u\|\|w\|\ .
\end{eqnarray*}
Recall that $\bx',\byy'$ are orthogonal to the first eigenvector of $A$.
Therefore, $\|\bx' A\|$ and $\|A\byy' \|$ are at most $\lambda_2 (A)$.
Then
$$
\mu_2(a)-\lambda_2(A)\leq \frac{a\lambda_2(A)}{\sqrt{n}-a}+
                          \frac{a\lambda_2(A)}{\sqrt{n}-a}+
                          \frac{a^2\lambda_1(A)}{(\sqrt{n}-a)^2}
                      =   \frac{2a\lambda_2(A)}{\sqrt{n}-a}+
                          \frac{a^2\lambda_1(A)}{(\sqrt{n}-a)^2}\,,
$$
and the lemma follows. \quad\quad$\Box$

\begin{lemma}\label{le45}
Let $A=(a_{ij})$ be an $n$ by $n$ real symmetric matrix.
Assume that $s$ and $X$ are positive numbers satisfying
 $\lambda_2 (A) \le s/2$ and
$\sum_{i=1}^n(\sum_{j=1}^na_{ij}-s)^2 \le X$. Then
there is a  number $c_1, |c_1| \le \sqrt{n}$ such that
$\|\bJ- c_1 \bv_1 \| \le 2\sqrt{X}/s$, where $\bv_1$ is a unit
eigenvector corresponding to $\lambda_1 (A)$.
\end{lemma}

\Proof.
Let $\bv_1, \dots, \bv_n$ be unit eigenvectors of
$A$, corresponding to the eigenvalues
$\lambda_1(A),\ldots,\lambda_n(A)$, respectively. Since these vectors
form an orthogonal basis of $\Rx^n$, we can express the vector $\bJ$
as their linear combination:
$$
\bJ =\sum_{i=1}^n c_i \bv_i\,,
$$
where $|c_1| \le \|\bJ\| = \sqrt{n}$. It is not too
difficult to check the following relations:
$$
\sum_{i=1}^n (\lambda_i(A) -s)^2 c_i^2 =\|(A-sI)\bJ\|^2 =
\sum_{i=1}^n(\sum_{j=1}^n a_{ij}-s)^2\ .
$$

By the assumptions of the lemma we get then
$$
X \ge \sum_{i=1}^n (\sum_{j=1}^n a_{ij}-s)^2 \ge
\sum_{i=2}^n (\lambda_i(A)-s)^2 c_i^2 \ge
\frac{s^2}{4} \sum_{i=2}^n c_i^2\ .
$$
Therefore,
$$
\|\bJ-c_1 \bv_1\|^2 =\sum_{i=2}^n c_i^2 \le \frac{4X}{s^2}\ ,
$$
as desired.\quad\quad $\Box$

\begin{lemma}\label{le46}
Let $a_{ij}$, $1\le j\le i\le n$ be independent random variables,
uniformly bounded by 1 in their absolute values. Assume that for $i>j$,
the $a_{ij}$ have a common expectation $p$. Define $a_{ij}=a_{ji}$ for
$j>i$. Then there exists an absolute constant $c>0$ so that for all
 $t>1$,
$$
Pr\left[\sum_{i=1}\left(\sum_{j=1}a_{ij}-np\right)^2\geq tn^2\right]
<e^{-ct^2}\ .
$$
\end{lemma}

\Proof For $1\le i\le n$, let $p_i=\Ex[a_{ii}]$. We define
$$
Y_i=(\sum_{j=1}^n a_{ij}-np)^2\,,
$$
then $Y=\sum_{i=1}^n(\sum_{j=1}^na_{ij}-np)^2=\sum_{i=1}^n Y_i$.

We first estimate from above the expectation of $Y_i$. Set
$b_{ij}=a_{ij}$ for all $j\ne i$, set also $b_{ii}=a_{ii}+p-p_i$. Then
$\Ex[b_{ij}]=p$ for all $1\le i,j\le n$. We obtain:
$$
Y_i=\left(\sum_{j=1}^n (b_{ij}-np)+(p_i-p)i\right)^2=
(\sum_{j=1}^n b_{ij}-np)^2 +2(p_i-p)\sum_{j=1}^n b_{ij}-np)+(p_i-p)^2\ .
$$
Recall that $b_{ij}$ are independent random variables with a common mean
$p$. Therefore
\begin{eqnarray*}
\Ex[Y_i]&=&\Ex[\sum_{j=1}^nb_{ij}-np)^2]+(p_i-p)^2\\
        &=& VAR[\sum_{j=1}^n b_{ij}]+(p_i-p)^2
         = \sum_{j=1}^n VAR[b_{ij}]+(p_i-p)^2\le n(1-p)+(p_i-p)^2\\
        &\le & n\,,
\end{eqnarray*}
for large enough $n$. This implies that $\Ex[Y]=\sum_{i=1}^n\Ex[Y_i]\le
n^2$.

Now, it is easy to see that for every $1\le j\le i\le n$, changing the
value of the random variable $a_{ij}$ can change the value of $Y$ by at
most $c_{ij}=O(n)$ (recall the assumption $|a_{ij}|\le 1$).
Then the so called "independent bounded difference inequality", proved
by applying the Azuma--Hoeffding martingale inequality
(see,. e.g., \cite{McD89}), asserts that for every $h>0$,
$$
Pr[Y-\Ex[Y]\geq h]\le \exp\{-h^2/2\sum_{1\le j\le i\le n} c_{ij}^2\}
                  \le \exp\{-h^2/O(n^4)\} \ .
$$
Substituting $h=(t-1)n^2$ and using the fact $\Ex[Y]\le n^2$, we get the
desired bound on the upper tail of $Y$.
\quad\quad$\Box$

\vskip 3mm

\noindent {\bf Proof of Lemma \ref{le42}.} Recall that by (\ref{fk2} we
have $\Ex[\lambda_1[A]=O(\sqrt{n})$. From the analysis of the case
$t \ge 10\sqrt{n}$ it follows then that $Pr[\lambda_2 \ge np/2] \le
e^{-c(np)^2} \le e^{-ct^2}$. Also, by Corollary \ref{co1}
$\Ex[\lambda_1(A)]=np +o(n)$.
Combined with our Theorem \ref{th1}, this implies that
$Pr[\lambda_1(A) \ge 2np] \le e^{-c(np)^2} \le e^{-ct^2}$. These two
facts, together with Lemma \ref{le46}, show that for that if
$t<10\sqrt{n}$, then with probability at least $1-e^{-ct^2}$,
the following three properties hold:
\begin{enumerate}
\item  $\sum_{i=1}^n (\sum_{j=1}^n a_{ij}-np)^2 \le  n^2t$;
\item  $\lambda_2 (A) \le np/2$\,;
\item  $\lambda_1 (A) \le 2np $\,.
\end{enumerate}

Assume that a matrix $A$ satisfies conditions 1, 2 and 3 above.
Applying (in this order) Lemma \ref{le45} with $X=n^2t$ and $s=np$ and
Lemma \ref{le44} with $a=2X^{1/2}/s=2t^{1/2}/p$,
we have that with probability at least $1-e^{-ct^2}$
$$
\mu_2 (A) -\left(1+ \frac{2a}{\sqrt{n}-a}\right)\lambda_2 (A) \le
\frac{a^2 \lambda_1(A)}{(\sqrt{n}-a)^2} \le
\frac{4X}{s^2}\, \frac{2np}{(\sqrt{n}-2\sqrt{X}/s)^2} \le
\frac{9t}{p}\ .
$$
Substituting the value of $a$, we get:
$$
Pr[\mu_2-\left(1+\frac{5\sqrt{t}}{\sqrt{n}p}\right)\lambda_2\ge
\frac{9t}{p}]<e^{-ct^2}\ .
$$
The proof is completed by rescaling, namely, by setting $t:=t/p$.
\quad\quad$\Box$

\vskip 3mm

\noindent{\bf Proof of Lemma \ref{le43}.} First notice that
$$\Ex [\mu_2] -\Ex [\lambda_2] =\Ex [\mu_2 -\lambda_2 ] \le
\int_{0}^{\infty} t Pr [\mu_2 -\lambda_2 \ge t] dt\ .
$$
Moreover,
$$
\int_{0}^{\infty} t Pr [\mu_2 -\lambda_2 \ge t] dt \le
\int_{0}^{K_p}t dt +
\int_{K_p}^{10 \sqrt{n}}t Pr[\mu_2 -\lambda_2 \ge t] dt +
\int_{10\sqrt{n}}^{\infty} t Pr [\mu_2 \ge t] dt\ .
$$
The first integral is clearly bounded by a constant depending on $p$r
 only.  By Lemma \ref{le41} and the fact that $\Ex(\mu_2)\le 3\sqrt{n}$,
$Pr [\mu_2 \ge t] \le e^{-ct^2}$ for $t\ge 10\sqrt{n}$. Thus
$\int_{10\sqrt{n}}^{\infty} t Pr [\mu_2 \ge t] dt \le
\int_{0}^{\infty} t e^{-ct^2} dt = O(1)$. To bound
the second integral, note that
\begin{eqnarray*}
\int_{K_p}^{10\sqrt{n}} t Pr[\mu_2 -\lambda_2 \ge t] dt &\le&
\int_{K_p}^{10\sqrt{n}} t Pr[\mu_2-\lambda_2\ge t/2+\ep_t\lambda_2] dt\\
&+& \int_{K_p}^{10\sqrt{n}} t Pr[\ep_t \lambda_2 \ge t/2] dt .
\end{eqnarray*}
By Lemma \ref{le42},
 $$
 \int_{K_p}^{10\sqrt{n}}t Pr[\mu_2 -\lambda_2\ge t/2+\ep_t\lambda_2]dt
\le \int_{0}^{10\sqrt{n}} t e^{-c_pt^2} dt = l_1\ ,
$$
where $l_1>0$ is a constant depending only on $p$.

On the other hand, we know that   $\ep_t\le b t^{1/2}(np)^{-1/2}$
for some constant $b$. Using the analysis of the case $t\ge 10\sqrt{n}$,
assume that $K_p > (30b/p)^2$; for any $K\le t\le 10\sqrt{n}$ we have:
$$
Pr[\ep_t \lambda_2 \ge t/2] \le Pr [\lambda_2\ge\frac{ t^{1/2}}{2b}
(np)^{1/2}] \le e^{-c_pt^2}\ .
$$
This implies that
$$
\int_{K_p}^{10\sqrt{n}} t Pr[\ep_t \lambda_2 \ge t/2] dt
\le  \int_{0}^{10\sqrt{n}} t e^{-c_pt^2} dt = l_2\ ,
 $$
where $l_2$ is a constant depending on $p$ only. This completes the
proof. \quad\quad$\Box$

\vskip 5mm

The proof for $\delta_2$ is similar. Instead of $\mu_2$, consider
$\mu_2'= \max_{\bx, \|\bx=1\|, \bx\bJ=0} \bx A\bx$. Again, using
Talagrand's inequality one can obtain a version of Lemma \ref{le41} for
$\mu'_2$. The rest of the proof is similar and  we omit the
details.
%\quad\quad $\Box$

\section{Concluding remarks}

\begin{itemize}
\item Unfortunately, we are unable to extend Theorem \ref{th2} to the
case when the expectation $p$ of the non-diagonal entries of a random
matrix $A_n$ is a function of $n$ and tends to zero as $n$ tends to
infinity, without imposing additional restrictions of the distribution
of entries. However, Theorem 2 can be extended in the following special
but important case: the diagonal entries of $A_n=(a_{ij})$ are all
zeroes, and the entries above the main diagonal are  i.i.d. Bernoulli
random variables with parameter $p=p(n)$, i.e., $Pr[a_{ij}=1]=p$ and
$Pr[a_{ij}=0]=1-p$ for all $1\leq i<j\leq n$. In this case the random
matrix $A_n$ can be identified with the adjacency matrix of a
{\em random graph} $G(n,p)$, and the eigenvalues of $A_n$ are the
eigenvalues of a random graph on $n$ vertices. Under these assumptions
we have the following result.

\begin{theorem}\label{theo3} There are positive constants $c$ and  $K$
  such that if $p= \omega (n^{-1})$ then for any $t > K$,

$$Pr [|\lambda_2 (A) - E[\lambda_2 (A)]| \ge t ] \le
e^{-ct^2}, $$

\noindent where $A$ is the adjacency matrix of $G(n,p)$. The same
result holds for $\delta_2 (A)$.
\end{theorem}

This theorem can be proved by repeating the arguments in the proof
of Theorem 2 under the new assumptions.  We have to make some
significant changes only in the proof of Lemma 4.6. The method of
bounded difference martingale (Azuma-Hoeffding's inequality) seems
not powerful enough to prove the the statement Lemma 4.6 when $p$
is decreasing in $n$, and we need to invoke a recent concentration
technique presented in \cite{Vu}. The details are omitted. Notice that
in many graph theoretic applications the eigenvalue $\lambda_2(A(G))$ is
of special importance as it reflects such graph properties as expansion,
convergence of a random walk to the stationary distribution etc.

\item Though we could show the tightness of our main result (Theorem
\ref{th1}) in some cases and for some values of the deviation parameter
$t$, it will be extremely interesting to reach a deeper understanding
of the tightness of Theorem \ref{th1} for the whole range of $t$ and for
some particular important distributions of the entries of $A$.

\item Theorem \ref{th1} is obtained under very general assumptions on
the distribution of the entries of a symmetric matrix $A$. Still, it
will be very desirable to generalize our result even further, in
particular, dropping or weakening the restrictive assumption about the
uniform boundness of the entries of $A$. This task however may require
completely different tools as the Talagrand inequality appears to be
suited for the case of bounded random variables.

\item Finally, it would be quite interesting to find further
applications of our concentration results in algorithmic problems on
graphs. The ability to compute the eigenvalues of a graph in polynomial
time combined with an understanding of potentially rich structural
information encoded by the eigenvalues can certainly provide a basis for
new algorithmic results exploiting eigenvalues of graphs and their
concentration.
\end{itemize}

\noindent{\bf Acknowledgment.} The authors are grateful to Zeev Rudnick
for his helpful comments.

\end{document}